\documentclass[11pt]{article}

\usepackage{amsmath}
\usepackage{amssymb}
\usepackage{extarrows}
\usepackage{graphicx}
 \usepackage{indentfirst}
\usepackage{hyperref}

\renewcommand{\baselinestretch}{1.3}
  \renewcommand{\arraystretch}{1.2}
\begin{document}

 \title{On the Weakness of Fully Homomorphic Encryption}

\author{Zhengjun Cao$^{1}$, \quad Lihua Liu$^{2}$}

  \footnotetext{\noindent $^1$Department of Mathematics, Shanghai University, Shanghai,
  China. \\
     $^2${Department of Mathematics, Shanghai Maritime University, Shanghai,  China. \, \textsf{liulh@shmtu.edu.cn} 
  }}

 \date{}\maketitle

\begin{quotation}
 \textbf{Abstract}. Fully homomorphic encryption (FHE) allows anyone to perform
 computations on encrypted data, despite not having
the secret decryption key.  Since the Gentry's work in 2009, the primitive has interested many researchers.
In this paper, we stress that any computations performed on encrypted data are constrained to the encrypted domain (finite fields or rings).
 This restriction makes the primitive useless for most computations involving common arithmetic expressions and relational expressions.
It is only applicable to the computations related to modular arithmetic.
We want to reaffirm that cryptography uses modular arithmetic a lot in order to  obscure and  dissipate the redundancies in a plaintext message, not to perform any numerical calculations. We think it might be an overstated claim that FHE is of great importance to  client-server computing or cloud computing.

 \end{quotation}

\section{Introduction}

Homomorphic encryption introduced  by
Rivest, Adleman and Dertouzos \cite{RAD78}  in 1978,   is a useful cryptographic primitive because it can translate
 an operation on the ciphertexts into an operation on the corresponding  plaintexts.
  The property is useful for some applications, such as e-voting, watermarking and secret sharing schemes. For example,  if an additively homomorphic
encryption is used in an e-voting scheme, one can obtain an encryption of the sum of all
ballots from their encryption. Consequently, it becomes possible that a single decryption will reveal the result
of the election. That is, it is unnecessary to decrypt all ciphertexts one by one.

A fully homomorphic encryption (FHE) is defined as a scheme which allows anyone to perform arbitrarily
 computations on encrypted data, despite not having
the secret decryption key.
 In 2009,  Gentry \cite{G09} proposed  a FHE scheme over ideal lattices, which is capable of evaluating some functions in the encrypted domain.
 Since then,  the primitive has interested many researchers.

\subsection{Related works}
 Homomorphic encryption schemes supporting either addition or multiplication operations (but not both) had been intensively studied, e.g.,  Goldwasser-Micali encryption \cite{GM82}, ElGamal encryption \cite{E84}, and Paillier encryption \cite{P99}. The Gentry encryption  \cite{G09} is  a fully homomorphic encryption
 scheme, which makes it possible to evaluate some functions in the encrypted domain.  After that, some new FHE schemes appeared.

  At Eurocrypt'10,  Gentry, Halevi and Vaikuntanathan \cite{GHV10} proposed a FHE scheme  based on the Learning With Error (LWE) problem. In 2010, van Dijk, et al. \cite{DGHV10} constructed a simple FHE scheme using only elementary modular arithmetic.
At Crypto'11,  a FHE scheme working over integers with shorter public keys and a FHE scheme based on ring-LWE were presented by  Coron et al. \cite{C11}, Brakerski and Vaikuntanathan \cite{BV11}, separately.  At FOCS'11, a  FHE scheme based on standard LWE by Brakerski and Vaikuntanathan \cite{BV12,BV14}, and a  FHE scheme
using depth-3 arithmetic circuits by Gentry and  Halevi \cite{GH11}, have interested many audiences. In 2012,  Brakerski, Gentry  and  Vaikuntanathan \cite{BGV12} designed a leveled FHE scheme without bootstrapping.  At Eurocrypt'13,  Cheon, et al. \cite{C13} investigated the problem of batching FHE schemes over integers.
In 2013,  Brakerski, Gentry and Halevi \cite{BGH13} discussed the problem of packing ciphertexts in LWE-based
homomorphic encryption.

In 2015, Castagnos and Laguillaumie \cite{CL15} proposed  a linearly homomorphic encryption scheme whose
security relies on the hardness of the decisional Diffie-Hellman problem. Recently, Cheon and   Kim \cite{CK15} introduced a hybrid homomorphic encryption
which combines public-key encryption and somewhat homomorphic encryption in order to reduce the storage requirements
for some applications.

 FHE makes it possible to enable secure storage and computation on the cloud. However, current homomorphic encryption
schemes are still inefficient. For example, key generation in Gentry's FHE scheme takes from 2.5 seconds to 2.2 hours \cite{GHE11}.
 A recent implementation required 36 hours for a homomorphic evaluation of AES \cite{GHS12}.  One of the most remarkable things about these implementations is that
the computations did not involve common arithmetic expressions and relational expressions.

\subsection{Our contributions}
 In this paper, we want to stress that any computations performed on encrypted data are constrained to the encrypted domain (finite fields or rings).
 This restriction makes the primitive useless for most computations involving common arithmetic expressions, logical expressions and relational expressions.
It is only applicable to the computations related to modular arithmetic.
Some researchers have neglected the differences between common arithmetic and modular arithmetic, and falsely claimed that FHE enables arbitrary computations on encrypted data.
We here reaffirm that cryptography uses modular arithmetic a lot in order to  obscure and  dissipate the redundancies in a plaintext message, not to perform any numerical calculations.

We revisit the Dijk-Gentry-Halevi-Vaikuntanathan FHE scheme \cite{DGHV10} and Nuida-Kurosawa FHE scheme \cite{NK15} under the client-server computing model.
The former encrypts bit by bit. The latter works over the encrypted domain $\mathbb{Z}_Q$, where $Q$ is a prime.
We find that in the Dijk-Gentry-Halevi-Vaikuntanathan scheme the server can not decide the carries by the encrypted data, and in the Nuida-Kurosawa scheme it is impossible to find an invertible transformation $\mathcal{T}$ from the real number set $\mathbb{R}$ to the field $\mathbb{Z}_Q$. Therefore, in both schemes the server can not return right values to the client even though the server is asked to help to evaluate the simple function $f(x, y)=x+y$.

In view of  the limitations mentioned above, we believe it might be a false claim that FHE is of great importance to cloud computing.
To the best of our knowledge, it is the first time to concretely  discuss FHE schemes under the client-server computing  model.

\section{The real goal of using modular arithmetic in cryptography}

Any calculation needs an describing expression, which consists of variables, constants and operators. There are three kinds of expressions: arithmetic expressions, logical expressions and relational expressions.   Arithmetic operators include addition $(+)$, substraction $(-)$, multiplication ($*$), division $(/)$, integer-division $(\setminus)$, modulus (Mod), and so on.

Like common arithmetic, modular arithmetic is commutative, associative, and distributive.
Suppose that $a, b$ are in the decrypted domain $\mathbb{Z}_p$ where $p$ is a prime, $E(\cdot)$ is a fully homomorphic encryption algorithm, and $D(\cdot)$ is the corresponding  decryption  algorithm. Then the following properties are obvious.
$$D(E(a)+E(b))=D(E(a+b))=a+b \mod p$$
$$D(E(a)\cdot E(b))=D(E(ab))=ab \mod p.$$
Generally,
$$ a+b \not=(a+b \mod p), \qquad ab\not= (ab \mod p)   $$
$$ a<b  \not\Longrightarrow E(a)<E(b), \qquad E(a)<E(b) \not\Longrightarrow  a<b$$

We here want to stress that cryptography uses modular arithmetic a lot, because it can obscure the relationship between the plaintext and the ciphertext, and dissipate the redundancy of the plaintext by spreading it out over the ciphertext. It is well known that confusion and diffusion are the two basic techniques for obscuring the redundancies in a plaintext message. They could frustrate attempts to study the ciphertext looking for redundancies and statistical patterns.
Practically speaking, the real goal of using modular arithmetic in cryptography is to \emph{obscure and  dissipate the redundancies in a plaintext message, not to perform any numerical calculations.}

To see this, we will have a close look at two typical  FHE schemes proposed by van Dijk et al. \cite{DGHV10},  Nuida and Kurosawa \cite{NK15}.
The former encrypts bit by bit. The encrypted domain for the latter  is $\mathbb{Z}_Q$, where $Q$ is a prime.

\section{Analysis of Dijk-Gentry-Halevi-Vaikuntanathan FHE scheme under the client-server computing  model}

\subsection{Description of Dijk-Gentry-Halevi-Vaikuntanathan scheme}

At Eurocrypt 2010, van Dijk et al. \cite{DGHV10} constructed an FHE scheme.
For convenience, we here only describe the symmetric version  of  the Dijk-Gentry-Halevi-Vaikuntanathan FHE scheme as follows.

\textbf{KeyGen}($\lambda$): For a security parameter $\lambda$, pick an odd number $p \in[2^{\lambda-1}, 2^{\lambda})$ and set it as the secret key.

\textbf{Encrypt}($p, m$): Given a bit $m\in\{0, 1\}$, compute the ciphertext as
$$  c = pq + 2r + m$$ where the integers $q, r$ are chosen at random in some other
prescribed intervals, such that $2r$ is smaller than $p/2$ in absolute value.

\textbf{Decrypt}($p, c$): $m=(c \mod p) \mod 2$.

\textbf{Additively homomorphic property (under the modulus)}: If $c_1 = pq_1 + 2r_1 + m_1$ and  $c_2 = pq_2 + 2r_2 + m_2$, then $m_1+m_2=(c_1+c_2\mod p)\mod 2$.

\textbf{Multiplicatively homomorphic property (under the modulus)}: If $c_1 = pq_1 + 2r_1 + m_1$ and  $c_2 = pq_2 + 2r_2 + m_2$, then $m_1\cdot m_2=(c_1 \cdot c_2\mod p)\mod 2$.

Notice that these homomorphic properties hold only on the condition that computations are constrained by the prescribed modulus $p, 2$.
This restriction makes the scheme impossible to deal with any numerical calculations without knowing the modulus.

\subsection{An example for Dijk-Gentry-Halevi-Vaikuntanathan scheme}
 Suppose that one client sets $p=7919$ as his secret key.
 He has two numbers  $a=5$, $b=3$, and wants a server to help him to compute  $c=a+b$.
 Now, he encrypts two numbers  $a$ and $b$ as follows (see Table 1).
\begin{center}
\begin{tabular}{l|ccc}
  \hline
$a=5$& \fbox{1}& \fbox{0}& \fbox{1} \\ \hline
 & $7919\times 1325+2\times 57+1 $ & $7919\times 3168+2\times 49+0 $ & $7919\times 5247+2\times 63+1 $ \\
&  \fbox{10492790} & \fbox{25087490} & \fbox{41551120} \\
  \hline \hline
 $b=3$&  & \fbox{1}& \fbox{1} \\ \hline
 &  & $7919\times 5538+2\times 85+1 $ & $7919\times 6214+2\times 74+1 $ \\
&  & \fbox{43855593} & \fbox{49208815} \\
  \hline
\end{tabular}\vspace*{3mm}

Table 1: Ciphertexts of 5 and 3 w.r.t. the secret key $7919$
\end{center}
The client sends two ciphertexts
$$ \underbrace{\fbox{10492790},\fbox{25087490},\fbox{41551120}}_x \
\mbox{and}\  \underbrace{\fbox{43855593},\fbox{49208815}}_y$$
to a server and asks the server to compute the function
$$f(x, y)=x+y.$$
 Hence, the server may return the values
$$  \fbox{10492790},\fbox{68943083},\fbox{90759935} $$
to the client.  Thus, the client decrypts the returned values as follows
$$(10492790 \mod p) \mod 2= 1,$$
$$(68943083 \mod p) \mod 2= 1,$$
$$(90759935 \mod p) \mod 2= 0,$$
and obtains the number $(110)_2=6$, not the right number $8$. See the following Table 2 for the process.

 \begin{center}
 \begin{tabular}{|lcl|}
   \hline
    Client  &   & Server \\ \hline
   Input: $p=7919$, & & $f(x, y)=x+y$\\
   \hspace*{10mm}  $a=5$, $b=3$   &   &   \\ \hline
      Encryption: $3 \rightarrow \underbrace{\fbox{43855593},\fbox{49208815}}_y $,&&\\
      $5\rightarrow \underbrace{\fbox{10492790},\fbox{25087490},\fbox{41551120}}_x$. && \\
  & $\xlongrightarrow{x, y}$    & \\
  & $\xlongleftarrow{\hat c}$ &  $ f(x, y)=\underbrace{\fbox{10492790},\fbox{68943083},\fbox{90759935}}_{\hat c}$\\
  Decryption: $\hat c \rightarrow \underbrace{\fbox{1},\fbox{1},\fbox{0}}_c$ &  & \\
    \hline
 \end{tabular}
 \vspace*{3mm}

Table 2: An example for the Dijk-Gentry-Halevi-Vaikuntanathan FHE scheme
 \end{center}

What is the problem with this process?
The returned values miss all carries because \emph{the server can not decide the carries by the encrypted data.}

\textbf{Remark 1}. One might argue that the client himself can construct a Boolean circuit which contains the carries and send the circuit to the server.
The argument is unreasonable because the client is assumed to be of weak computational capability.  If the client can construct such a Boolean circuit, then
he can directly evaluate the circuit, instead of asking a server to help him to evaluate it.

 \section{Analysis of Nuida-Kurosawa FHE scheme under the client-server computing  model}

 In the Dijk-Gentry-Halevi-Vaikuntanathan FHE scheme, the message space is $\mathbb{Z}_2$.
The scheme is very inefficient because it has to generate 256 or more bits in order to mask one bit.
 At Eurocrypt 2015, Nuida and Kurosawa \cite{NK15} extended the scheme to the message space $\mathbb{Z}_Q$ where $Q$ is any prime. We here only  describe the symmetric version of Nuida-Kurosawa FHE scheme as follows.

 \subsection{Description of Nuida-Kurosawa FHE scheme}

\textbf{KeyGen}($\lambda$): For a security parameter $\lambda$, pick an odd number $p \in[2^{\lambda-1}, 2^{\lambda})$ and a prime $Q$. Set $p$ as the secret key ($Q$ is published).

\textbf{Encrypt}($p, m$): Given a message $m\in \mathbb{Z}_Q$, compute the ciphertext as
$$  c = pq + Qr + m$$ where the integers $q, r$ are chosen at random in some other
prescribed intervals, such that $Qr$ is smaller than $p/2$ in absolute value.

\textbf{Decrypt}($p, c$): $m=(c \mod p) \mod Q$.

\textbf{Additively homomorphic property (under the modulus)}: If $c_1 = pq_1 + Qr_1 + m_1$ and  $c_2 = pq_2 + Qr_2 + m_2$, then $m_1+m_2=(c_1+c_2\mod p)\mod Q$.

\textbf{Multiplicatively homomorphic property (under the modulus)}: If $c_1 = pq_1 + Qr_1 + m_1$ and  $c_2 = pq_2 + Qr_2 + m_2$, then $m_1\cdot m_2=(c_1 \cdot c_2\mod p)\mod Q$.

\subsection{An example for Nuida-Kurosawa FHE scheme}
 Suppose that one client sets $p=22801763489$ as his secret key and sets $Q=15485863$.
 He has two numbers  $a=0.1$, $b=2.3$, and wants a server to help him to compute  $c=a+b$.

 First, he has to transform $a=0.1$, $b=2.3$ into integers $\bar a, \bar b$ such that $\bar a, \bar b\in \mathbb{Z}_Q$. Denote the transformation by $\mathcal{T}$.
 Second, he encrypts $\bar a, \bar b$ and obtains the corresponding ciphertexts $\hat a, \hat b$. Third, he sends $\hat a, \hat b$ to
 a server. The server then takes $\hat a, \hat b$ as the inputs of the function $f(x, y)=x+y$. Finally, the server returns $\hat c=f(\hat a, \hat b)$ to the client.
 See the following Table 3 for the process.

 \begin{center}
 \begin{tabular}{|lcl|}
   \hline
    Client  &   & Server \\
   Input: $p=22801763489$, $Q=15485863$;& & $f(x, y)=x+y$\\
   \hspace*{10mm}  $a=0.1$, $b=2.3$   &   &   \\ \hline
      Transformation $\mathcal{T}$: $ a\rightarrow \bar a$, $ b\rightarrow  \bar b$. & & \\
     \hspace*{14mm}  such that $\bar a, \bar b\in \mathbb{Z}_Q$. & & \\
   Encryption: $\bar a\rightarrow\hat a$, $\bar b\rightarrow  \hat b$.
  & $\xlongrightarrow{\hat a, \hat b} $    & \\
  & $\xlongleftarrow{\hat c}$ & Computation $ f(\hat a,\hat b)\rightarrow \hat c$\\
  Decryption: $\hat c \rightarrow \bar  c$ &  & \\
  Inverse Transformation $\mathcal{T}^{-1}$: $\bar c \rightarrow c. $ & & \\
   \hline
 \end{tabular}
 \vspace*{3mm}

Table 3:   An example for the Nuida-Kurosawa FHE scheme
 \end{center}

What is the problem with this process?  \emph{It is impossible to find an invertible transformation $\mathcal{T}$ from the
real number set $\mathbb{R}$ to the field $\mathbb{Z}_Q$.}

Note that most  encryption algorithms must run over some finite field or ring. One has to transform all inputting characters into integers in the field or ring.
That means an invertible encoding algorithm is necessary for any encryption scheme.

This condition is easily satisfied if all inputting characters are indeed viewed as characters. But when some inputting characters are viewed as real numbers and they are used for some arithmetic computations, it is impossible to find such an invertible encoding algorithm that maps any real number to an integer in a prescribed field or ring.

\begin{center}
\begin{tabular}{|cc|cc|}
  \hline
   character & ASCII code & character & ASCII code \\ \hline
  0 & 48 &  6 & 54 \\
 1 & 49   &   7 & 55  \\
 2 & 50  & 8 & 56 \\
 3 & 51  &  9 & 57  \\
 4 & 52  &  $\cdot$ & 250 \\
  5 & 53   & & \\
  \hline
\end{tabular}\end{center}

We here describe a possible encryption-decryption process for the real numbers $0.1$ and $2.3$. The ASCII coding method will map $0.1, 2.3$ to two integers in the field $\mathbb{Z}_{15485863}$.

\begin{center}
\begin{tabular}{|c|}
  \hline
  $a=0.1\xlongrightarrow{\mbox{ASCII}} \fbox{48}\fbox{250}\fbox{49}\xlongrightarrow{\mathcal{T}}\bar a=48\times 256^2+250\times 256+49=3209777\xlongrightarrow{q=3215964,r=13} $\\
$\hat a= 73329650721664392\xlongrightarrow{\mod p, \mod Q} \bar a=3209777
\xlongrightarrow{\mathcal{T}^{-1}}\fbox{48}\fbox{250}\fbox{49}\xlongrightarrow{\mbox{ASCII}} 0.1 $ \\ \hline
$b=2.3\xlongrightarrow{\mbox{ASCII}} \fbox{50}\fbox{250}\fbox{51}\xlongrightarrow{\mathcal{T}}\bar b=50\times 256^2+250\times 256+51=3340851\xlongrightarrow{q=6490231,r=9} $\\
$\hat b= 147988712393689577\xlongrightarrow{\mod p, \mod Q}\bar b = 3340851
\xlongrightarrow{\mathcal{T}^{-1}}\fbox{50}\fbox{250}\fbox{51}\xlongrightarrow{\mbox{ASCII}} 2.3 $
    \\   \hline
\end{tabular}\end{center}

If a server performs the operator of addition on the encrypted data, $\hat a, \hat b $, then it gives
$$ \hat c= \hat a+\hat b = 73329650721664392+147988712393689577= 221318363115353969.$$
The server returns the value  to the client. The client will obtain
$$\bar c= (221318363115353969 \mod p) \mod Q=6550628.$$
Notice that
$$6550628=99\times 256^2+244\times 256+100\xlongrightarrow{\mathcal{T}^{-1}} \fbox{99}\fbox{244}\fbox{100}.$$
It does not correspond to the wanted number $2.4$ when ASCII coding method is used.

\section{FHE is not applicable to client-server computing}

Cloud computing refers to the practice of transferring computer services such as computation or data storage to other redundant offsite locations available on the Internet, which allows application software to be operated using internet-enabled devices. It benefits one from the existing technologies and paradigms, even though he is short of deep knowledge about or expertise with them. The cloud aims to cut costs, and helps the users focus on their core business instead of being impeded by IT obstacles. Usually, cloud computing adopts the client-server business model.

What computations do you want to outsource privately?  Backup your phone's contacts directory to the cloud?  Ask the cloud to solve a mathematic problem in your homework? Do a private web search? $\cdots$. It seems obvious that the daily computational tasks are rarely constrained by some prescribed modulus. Moreover,
the client-server computing model can not deal with relational expressions which are defined over plain data, not over encrypted data. This is because
$$ a<b  \not\Longrightarrow E(a)<E(b), \qquad E(a)<E(b) \not\Longrightarrow  a<b.$$
In view of this weakness of FHE and the flaws of two typical schemes mentioned above, we think,  FHE is not applicable to cloud computing.

\textbf{Remark 2}.  The problem that what computations are worth delegating privately by individuals and companies to untrusted devices or servers remains untouched. We think the cloud computing community has not yet found a good for-profit model convincing individuals to pay for this or that computational service.

\section{Conclusion}

We reaffirm the role of modular arithmetic in modern cryptography and
show that FHE is not applicable to cloud computing because any FHE scheme does work over some encrypted domains. When two decrypted number are added, one cannot decide the carries
 without knowing the secret decryption key. Moreover, there is no an invertible transformation from the
real number set to the encrypted domain which makes it impossible to tackle numerical calculations.
We think the primitive of FHE might be of little importance to client-server computing.

\end{document}